# EEG Emotion Recognition Through Deep Learning

Roman Dolgopolyi, The American College of Greece, r.dolgopolyi@acg.edu,
https://orcid.org/0009-0002-9884-8386

Antonis Chatzipanagiotou, Independent Researcher, chatzipanagiotou.antonios@outlook.com,
https://orcid.org/0009-0005-7460-0062, Correspondence Author

**Abstract.** An advanced emotion classification model was developed using a CNN-Transformer architecture for emotion recognition from EEG brain wave signals, effectively distinguishing among three emotional states, positive, neutral, and negative. The model achieved a testing accuracy of 91%, outperforming traditional models such as SVM, DNN, and Logistic Regression. Training was conducted on a custom dataset created by merging data from SEED, SEED-FRA, and SEED-GER repositories, comprising 1,455 samples with EEG recordings labeled according to emotional states. The combined dataset represents one of the largest and most culturally diverse collections available. Additionally, the model allows for the reduction of the requirements of the EEG apparatus, by leveraging only 5 electrodes of the 62. This reduction demonstrates the feasibility of deploying a more affordable, consumer-grade EEG headset, thereby enabling accessible, at-home use, while also requiring less computational power. This advancement sets the groundwork for future exploration into mood changes induced by media content consumption, an area that remains underresearched. Integration into medical, wellness, and home-health platforms could enable continuous, passive emotional monitoring, particularly beneficial in clinical or caregiving settings where traditional behavioral cues, such as facial expressions or vocal tone, are diminished, restricted, or difficult to interpret, thus potentially transforming mental health diagnostics and interventions. Moreover, future studies can explore this further by leveraging this framework to develop con-tent recommendation algorithms based, not only on user retention, but also on emotional state derived from the EEG signal, personality traits, and user satisfaction. The model of this paper is a step towards a future personalized approach aiming to enhance mental well-being alongside traditional engage-ment metrics, which the authors are preparing a future study for. Following the introduction, section 2 describes previous works, utilizing similar deep learning methodologies. In the next section, the background theory of EEG is provided in order to examine thereafter, Emotion Recognition methods. In section 4 the preprocessing steps of the EEG data, that were used for this paper are described, with the architecture of the implemented deep learning model. Section 5 outlines the results of the analysis and compares the prediction accuracy with the results of pre-existing models. Lastly sections 6-8 provide the discussion, limitations and future directions.

# 1 Introduction

Many branches of health care are readily utilizing deep learning frameworks in professional environments and consumer-grade products and have allowed the advancement of science and the improvement of the consumer experience [1,2]. How-ever, in Neuroscience a lack of implementing Deep Learning methodologies seems to exist [3,4].

To this end this paper addresses the aforementioned gap by developing a Transformer Convolutional Neural Network, informed by the Electroencephalograph (EEG) brain wave signals, to classify different Emotional States (positive, neutral, negative). A multicultural dataset was utilized to train the model by merging the SEED [5,6], SEED-FRA [7,8] and SEED-GER [7,8] datasets, achieving one of the largest (1,455 samples) and one of the most diverse (Chinese, French and German population) EEG Emotion Recognition datasets [9]. This dataset included segments of Brain Wave signals and their subsequent labels of Emotion. This was achieved by showing clips validated to induce different emotional states and recording the brain waves via EEG [8].

Another novel aspect of this research endeavor is the showcasing that the number of required EEG electrodes, for emotion recognition, could be significantly reduced while achieving high accuracy. In this study the number of electrodes was reduced from 62 to 5. This selection was made to allow for emotion recognition, with EEG headsets that are portable and have a low barrier of entry in terms of ease of use and affordability. Such a readily available headset is the EMOTIV Insight, which informed the particular selection of 5 electrodes and their specific locations on the scalp [10]. Another important aspect which was considered was the proven high accuracy of the Insight and reliability in non-laboratory, noisy in electrical signals, everyday conditions [11–13].

In general, this research paper validates the applicability of EEG headsets and deep learning models in various fields and environments, which previously required costly non-portable equipment, a high level of specialization, high computational power and ideal conditions, which will be further explored in the Discussion section.

# 2 Related Work

## 2.1 Emotion Recognition

Early approaches in EEG-based emotion recognition primarily relied on classical machine learning algorithms such as Support Vector Machines (SVMs) and Random Forests, utilizing handcrafted features like Power Spectral Density (PSD) and Differential Entropy (DE) [14,15]. However, these methods often lacked generalizability and required domain expertise to manually engineer relevant features [16].

Recent developments have seen a paradigm shift towards deep learning models, which can automatically learn hierarchical feature representations from raw or minimally preprocessed EEG data. Convolutional Neural Networks (CNNs) have been extensively employed to capture spatial patterns in EEG signals by treating multi-channel EEG as structured input, such as 2D images or spatial graphs [16]. Recurrent Neural Networks (RNNs), particularly Long Short-Term Memory (LSTM) networks, have been utilized to model the temporal dynamics of brain activity [17]. Hybrid models, such as CNN-LSTM architectures, have demonstrated promising results by leveraging both spatial and temporal aspects of EEG data [17].

Transformer-based models, originally introduced for natural language processing, have recently been adapted for EEG analysis due to their capacity to capture long-range dependencies and global attention mechanisms [18]. Though still relatively novel in the EEG domain, these architectures are showing competitive performance compared to traditional CNN and RNN-based models [19].

### 2.2 Emotion Recognition EEG Datasets

The datasets most frequently used for emotion recognition research include the DEAP [20], DREAMER [21], and SEED [22] datasets. Among them, the SEED family of datasets (SEED, SEED-FRA, SEED-GER) stands out due to its large sample size, high-quality recordings, and well-validated emotion induction protocols [5,23] The inclusion of multilingual and multicultural samples in SEED-FRA and SEED-GER allows for a deeper exploration of how cultural context influences emotion-related EEG patterns [7]. Despite the richness of these datasets, few studies have attempted to merge them to increase both sample size and diversity, highlighting a gap this paper aims to fill.

### 2.3 Real-time Flexible EEG Headsets

Another underexplored area is the adaptation of EEG-based emotion recognition models for real-time or consumer-grade applications. Most state-of-the-art results rely on dense EEG systems with up to 62 electrodes, which are impractical for everyday use [24]. Recent research has begun exploring reduced-channel models that retain classification performance while using fewer electrodes, often guided by neuroanatomical relevance or channel selection algorithms [25,26]. Devices such as the EMOTIV Insight, which provide only five channels (Fig. 2), are beginning to be validated for real-world applications outside laboratory settings [27]. However, few studies combine such low-density EEG with advanced deep learning models in cross-cultural contexts, a direction this paper addresses.

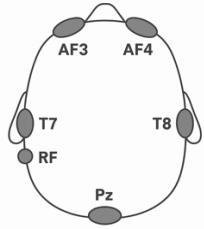

**Fig. 1.** Electrode **placement of the EMOTIV Insight EEG headset.** The five active electrodes, AF3 (Anterior Frontal 3), AF4 (Anterior Frontal 4), T7 (Temporal 7), T8 (Temporal 8), and Pz (Parietal midline), are positioned according to the international 10–20 system. An additional reference electrode (RF), placed behind the left ear, serving as a stable baseline for voltage comparison, allowing the EEG system to measure relative changes in brain electrical activity. The reference electrode is essential for minimizing noise and improving signal accuracy across all recording channels.

### 2.4 Applications of EEG-Based Emotion Recognition

Lastly, integration of EEG emotion recognition with intelligent systems like media recommender engines remains an emerging topic. While some frameworks use facial expressions or speech to recommend content, EEG-based recommendation systems are rare [20,28] The application of deep learning in mapping emotional states to content preferences is promising, particularly in systems that adapt dynamically to users without requiring manual input or self-reports [29,30]. EEG-based emotion recognition has progressed from laboratory-based experimentation to broader real-world applications, including consumer products, neuromarketing, and mental health technologies. As portable and affordable EEG devices become more widespread, the integration of emotion recognition into user-centered systems has gained traction.

**Consumer Domain Applications.**

In the consumer domain, lightweight headsets like the EMOTIV Insight have enabled non-invasive emotional monitoring in real time [11]. These devices open up opportunities for affect-aware recommendation systems, which adapt content, such as movies, music, or interactive experiences, based on users' brain-derived emotional states [12]. Unlike facial or voice recognition systems, EEG enables a passive and less intrusive approach, offering increased reliability in environments where expressive behaviors may be suppressed or misleading [20,28]. Deep learning models enhance this by detecting nuanced patterns in brain signals without the need for self-reporting, enabling seamless integration into smart home systems, streaming platforms, or wearable technologies [16,17].

**Neuromarketing Applications.**

In neuromarketing, EEG emotion recognition allows companies to analyze consumers' subconscious reactions to advertisements, packaging, or product designs [31]. The capacity to monitor emotional engagement provides insights that traditional surveys or focus groups often miss. Previous EEG headsets used in such studies were often bulky, wired, and uncomfortable for long durations, leading to participant fatigue and unnatural movement restrictions, which could bias results [32].

With the emergence of lightweight, wireless headsets, participants can now engage more freely and naturally with content, improving ecological validity and enhancing the overall reliability of neuromarketing assessments [33]. Brands leverage EEG-informed metrics, such as arousal and valence, to fine-tune marketing strategies and enhance user experience design. Studies have shown that EEG-based measurements correlate with brand recall, purchase intention, and attention patterns, making EEG a valuable tool in market research [34,35].

**Medical Applications**.

From a medical and mental health perspective, EEG emotion recognition supports the development of digital therapeutic tools for emotional regulation. For instance, emotion-aware interfaces can help users with anxiety, depression, or PTSD by providing emotionally congruent or calming content [36]. These systems function as passive emotional support tools that complement traditional therapies and digital cognitive behavioral interventions [37]. Mood-lifting content may also be employed as a form of digital therapeutic intervention, which may offer passive emotional regulation support in daily life, by complementing traditional treatments or improve mental well-being in non-clinical populations [36,38]. Furthermore, EEG-based neurofeedback systems use real-time emotional monitoring to guide users in managing their mental states, showing promise in reducing stress and improving emotional resilience [39,40]. Applications such as mindfulness training, emotion regulation support, and early detection of mood disorders have all shown potential when using EEG in clinical or semi-structured environments [38]. This paper contributes to this niche by proposing an emotion-aware recommendation engine that utilizes EEG input processed by a Transformer-CNN model to suggest relevant content from large movie/series datasets such as IMDb.

Together, these domains highlight the versatility of EEG-based emotion recognition. By combining consumer-grade EEG with deep learning, real-time emotional understanding is becoming accessible, scalable, and impactful across diverse applications.

## 3 Key Concepts

### 3.1 1.1 Basics of Brain Waves and EEG

Neurons communicate through both electrical and chemical signals [41]. When a neuron fires, it generates an action potential—an electrical pulse that travels along the axon to release chemical messengers called neurotransmitters [42]. These neurotransmitters cross synapses to activate or inhibit neighboring neurons, creating electrical changes in their membranes known as postsynaptic potentials [43]. While EEG does not measure action potentials directly, it records the summed postsynaptic potentials from thousands of neurons [43]. In simpler terms, EEG captures the echo of neuron-to-neuron conversations, rather than the spike of a single voice.

Electroencephalography (EEG) is a non-invasive method for recording the brain's electrical activity by detecting voltage fluctuations on the scalp caused by neuronal activity within the brain [41]. In simpler terms, EEG is like placing sensors on the head to listen to the brain's natural electrical signals. These signals are mainly generated by pyramidal neurons in the cerebral cortex, which are organized in columns and have long branches (dendrites) pointing toward the surface of the brain [42]. Because of their shape and alignment, when many of these neurons are activated together, they produce small electrical currents that can be detected on the scalp [3]. EEG signals travel through what is called a "volume conductor," meaning they pass through brain tissue, cerebrospinal fluid, the skull, and the skin before reaching the electrodes [4]. As such it is important to consider the possible limitations or distortions, that should be accounted for, as the electrical ripples are traveling through several layers before reaching the surface and recorded [4]. EEG setups use electrode caps placed on the scalp to detect these voltages across various brain regions (Fig. 1-6) [44]. Standardized placement systems (e.g., the 10–20 system) ensure consistent spatial coverage across participants [45].

Different brain wave frequencies reflect different mental and emotional states. For example, low-frequency delta waves (0.5–4 Hz) are prominent during deep sleep, while higher-frequency beta (13–30 Hz) and gamma (>30 Hz) waves are linked to focused attention and problem-solving [41]. Alpha waves (8–13 Hz) are common during relaxed, wakeful states and tend to decrease when someone becomes mentally active [46]. EEG captures both rhythmic and irregular patterns, and the exact shape and frequency of these patterns help neuroscientists and AI models understand brain states [47]. Due to its very high temporal resolution, EEG is ideal for tracking emotional changes as they happen [5]. This makes it a powerful tool for real-time applications, including emotion-aware systems that adapt to the user's current mental state.

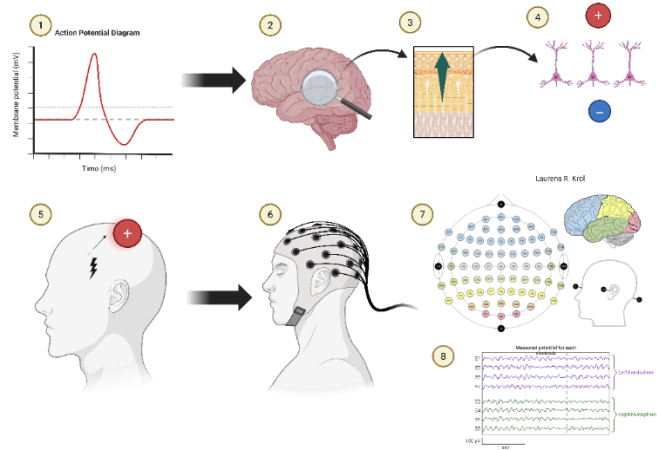

**Fig. 2.** Overview of EEG Signal Generation and Acquisition Process. (1) Neurons communicate through electrical signals known as action potentials. (2) Collective electrical changes, rather than individual spikes, are what EEG measures. (3) EEG signals primarily originate from pyramidal neurons located in the cerebral cortex. These neurons are aligned in vertical columns with their dendrites oriented toward the scalp, which allows for the summation of electrical activity to be detectable at the surface. (4) The synchronous activation of many pyramidal neurons produces summed postsynaptic potentials, generating strong voltage fluctuations. (5) These electrical signals pass through the brain tissue, cerebrospinal fluid, skull, and skin—collectively known as the volume conductor—before reaching the scalp. (6) Electrodes are non-invasively placed on the scalp to record voltage fluctuations. The montage shown demonstrates how the EEG cap detects brain activity from multiple scalp locations simultaneously. (7) Standard electrode placement systems (e.g., the 10–20 system) ensure consistent and reliable localization of brain signals. Electrodes are placed at intervals of 10% or 20% of these distances across four directions. The figure highlights various cortical regions targeted by EEG electrodes and the hemispheric layout. (8) The recorded output consists of waveforms representing voltage over time.

### 3.2 Emotion Recognition and Feature Extraction

As previously mentioned, motional states influence how brain waves behave— for instance, some emotions cause certain frequencies to become active [48]. This means that emotions can be inferred by analyzing EEG signals. To do this effectively, raw EEG data must be cleaned-preprocessed and thereafter analyzed [43]. Common noise sources like eye movements or muscle activity are first removed [49]. Then, features are extracted from different frequency bands. One widely used feature is Differential Entropy (DE), which measures how much variability exists in a given EEG signal segment [20]. DE is a reliable measurement in classifying multiple emotions because it captures how stable or chaotic the signal is and contrasts one signal to the distribution of all the signals, with each signal having a label (e.g. an emotional state) [50]. Thus, attributing the given signal to the point in the distribution that is the most similar to it and subsequently classifying it with that label that matches the signals in that point of the distribution [51].

Once these features are extracted, machine learning models are used to classify them. These models include traditional approaches like Support Vector Machines (SVMs) [9] and more advanced methods like Convolutional Neural Networks (CNNs), which can learn directly from raw or filtered EEG data [52]. Deep learning models in particular are good at identifying patterns across space and time in EEG data, making them especially useful for emotion recognition systems that need to generalize across users [53].

### 3.3 Cultural Variability in Emotion

Emotion expression is not solely a biological response but is heavily shaped by cultural norms, values, and social expectations, with variations documented across individualistic and collectivist societies [54]. Although basic emotional expressions such as happiness and anger are often cited as universal, Ekman and Friesen demonstrated that display rules, the culturally-specific norms about the appropriateness of emotion expression, modify how these emotions are exhibited across cultures [55]. Studies examining distinct emotions across countries found that while there were universal elements, the cultural context significantly altered the frequency and intensity of each emotion's appraisal [56–58]. For example, the emotion of pride may be appraised positively in individualistic cultures but viewed negatively in collectivist cultures where humility is emphasized [58].

Perception and recognition of emotional expressions are also subject to cultural differences [59]. To this extent differences in the brain responses across participants have also been observed, validating further the above claims [60].

Cultural orientation profoundly influences emotion regulation strategies. In collectivist cultures, emotional suppression is commonly practiced to avoid disrupting group cohesion, whereas individualistic cultures promote emotional expressiveness as a sign of authenticity [61].

In EEG-based emotion recognition research, it is also essential to consider how cultural differences may influence the neural correlates of emotional processing. Studies have shown that EEG features such as power spectral density (PSD), differential entropy (DE), and event-related potentials (ERPs) can vary depending on cultural context and emotional norms [62]. For example, cultural differences in frontal alpha asymmetry have been reported, suggesting distinct patterns of approach-avoidance motivation linked to culturally preferred emotion regulation strategies [63]. Additionally, emotion elicitation paradigms using culturally relevant stimuli yield different EEG signatures, indicating the necessity of culturally tailored protocols for accurate emotion classification [64].

These findings imply that EEG emotion recognition systems must incorporate diverse cultural data to ensure robustness and generalizability across populations [65].

These cultural disparities have practical implications for cross-cultural interactions and technology design. Misinterpretations of emotional expressions can arise in multicultural environments due to differing cultural norms [20]. This becomes especially relevant in the development of AI-based emotion recognition systems, which may produce biased outputs if trained on culturally homogeneous datasets [66,67]. As noted by Zheng et al., integrating multicultural datasets and developing adaptive algorithms is essential to ensure accurate and inclusive emotion recognition models [68].

### 3.4 EEG-based AI Media Recommendation

One understudied application of EEG emotion recognition is using it to recommend media, like movies or TV shows, based on the viewer's emotional state in real time [44]. Instead of asking users what they want to watch, the system detects their emotional state and suggests appropriate content. For example, a person who appears bored based on their EEG patterns might receive recommendations for more stimulating shows, while someone who appears stressed might be offered calming content. This mapping between detected emotions and media is often managed using tools like knowledge graphs or semantic embeddings, which relate content tags (like "thrilling" or "romantic") to specific emotional states [39]. Additionally, individual personality traits may influence recommendation preferences [69]. For example, individuals scoring high on conscientiousness may prefer recommendations that contrast their current emotional state, such as suggesting an uplifting movie during sadness to remind them of resilience and the transience of emotions [69]. On the other hand, individuals high in empathy might prefer media that aligns with their detected emotion, as it allows them to process and relate deeply to emotional narratives [70].

These systems rely solely on brain signals, allowing for hands-free and without the need and or unreliability of self-report, interaction, and making them especially valuable in accessibility, healthcare, research and immersive environments [71,72]. In addition, combining EEG with other data such as eye tracking or physiological signals can enhance accuracy, in research context but, EEG and deep learning models alone have proven to be highly accurate and practical, allowing for a consumer-grade approach, that does not need support from more data and can allow for the unsterile natural environments that apply outside of the laboratories [52]. AI systems represent a new generation of interfaces that adapt in real time to users, forming part of a broader movement toward brain-computer interfaces (BCIs) that close the loop between sensing and responding [53].

## 4 Methodology

### 4.1 Multicultural EEG Corpus Construction

The SEED, SEED-FRA and SEED-GER repositories —all licensed for academic research use—were concatenated to form a single multicultural dataset acquired with an identical 62-electrode EEG system under comparable laboratory conditions, ensuring protocol compatibility [5,7,51]. SEED supplies 675 Chinese trials that are evenly divided into negative, neutral and positive labels (225 negative, 225 neutral and 225 positive). SEED-FRA contributes 480 French trials (144 negative, 168 neutral and 168 positive), and SEED-GER adds 300 German trials (80 negative, 120 neutral and 100 positive). After integrity checks, the merged dataset contains 1 455 trials (449 negative, 513 neutral and 493 positive) showing a class-imbalance ratio of 13.2 %, which is low enough to mitigate bias during training.

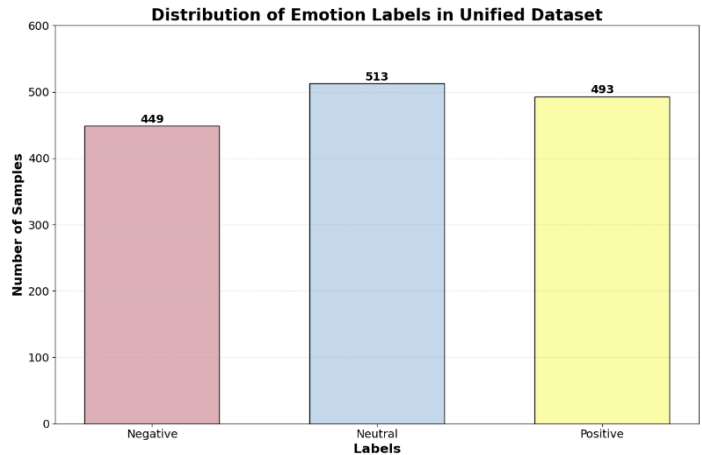

**Fig. 3.** Bar chart showing label distribution among obtained multicultural dataset.

### 4.2 Signal Pre-processing

Every 200 Hz, 62-electrode recording was reduced to the 5-electrode recording from the electrodes of the research interest (*AF3, AF4, T7, T8* and *Pz*), producing an $S \times 5$ matrix per trial, where $S$ is the number of samples. Each signal was then segmented into one-second windows and subjected to band-pass filtering to extract the *delta* (0.5–4 Hz), *theta* (4–8 Hz), *alpha* (8–13 Hz), *beta* (13–30 Hz), and *gamma* (30–45 Hz) frequency bands, which correspond to well-established classes of brainwave activity [73]. The choice of the one-second window length was motivated by its widespread adoption in prior EEG-based affective computing studies, providing a proven balance between temporal resolution and feature stability [74]. Band-specific variance in every epoch was converted to Differential Entropy (DE), generating a tensor of shape $T \times 5 \times 5$, where $T$ represents the number of one-second time windows, the second axis corresponds to the five frequency bands (*delta, theta, alpha, beta, gamma*), and the third axis denotes the five selected EEG channels (*AF3, AF4, T7, T8, Pz*).

### 4.3 Data-Partition Strategy

The dataset was randomly partitioned into three subsets: 70% for training (1,018 trials), 15% for validation (219 trials), and 15% for testing (218 trials). Because individual participants appear in more than one subset, the division is intentionally subject-dependent. This design maintains a sufficiently large training sample despite the limited number of volunteers per culture and reflects real-world deployment scenarios where a system can be recalibrated for each user.

### 4.4 Model Architecture and Hyper-parameter Optimization

To effectively learn both local and global patterns in the EEG data, a transformer-based CNN architecture was employed. The convolutional layers were used to extract localized spatio-spectral features, while the transformer's multi-head self-attention mechanism enabled the modeling of long-range temporal dependencies (Fig. 4).

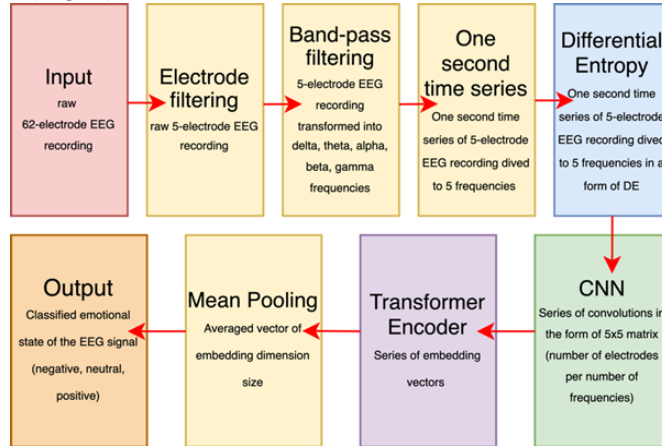

**Fig. 4.** Diagram that shows the pipeline of the considered model architecture.

To identify the most effective configuration of this architecture, nine hyper-parameters were systematically explored. A total of 100 configurations were sampled uniformly at random from a 1,024-point search space and each was proxy-trained for fifteen epochs. The configuration that achieved the highest validation accuracy was selected for further training (Table 1).

**Table 1.** that lists all hyper-parameters with their considered values for tunning process.

| Hyper-parameter | Search space |
|---|---|
| CNN output channels | {8, 16} |
| Kernel size | {3, 5} |
| Transformer layers | {2, 4} |
| Hidden units | {128, 256} |
| Attention heads | {4, 8} |
| Embedding dimension | {32, 64} |
| Batch size | {8, 16, 32} |
| Drop-out | {0.1, 0.3} |
| Learning rate | $\{1 \times 10^{-3}, 5 \times 10^{-4}\}$ |

## 5 Results

### 5.1 Training Protocol and Optimal Architecture

The optimal configuration—featuring 8 convolutional output channels with a 5-sample kernel, 4 transformer layers with 128 hidden units and 4 attention heads, a 64-dimensional embedding, batch size of 8, a dropout rate of 0.1, and a learning rate of $5 \times 10^{-4}$—was retrained for 100 epochs with Adam optimizer and categorical cross-entropy loss. Validation accuracy was monitored throughout, and the model state achieving the highest score of 0.9224 at epoch 89 was preserved for final testing. As shown in the plot, both training and validation accuracy improved steadily over time, with minimal divergence between the two curves. This indicates effective generalization and no signs of overfitting (Fig. 6).

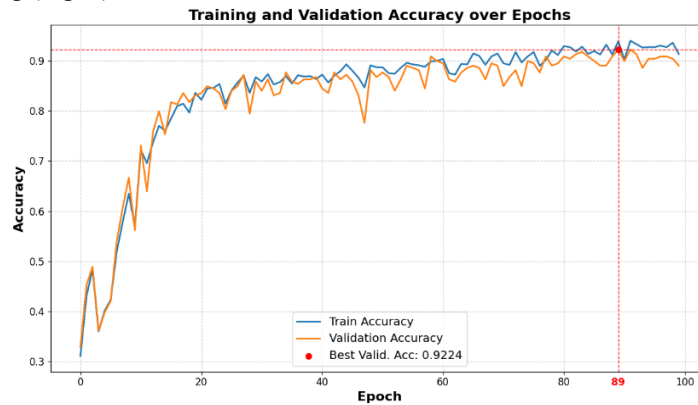

**Fig. 6.** Figure showing two trends of model prediction accuracy on train and validation datasets in relation to the number of epochs used for training, while highlighting the epoch at what the model reached its highest prediction accuracy on validation dataset.

## 5.2 Test-set Performance

When the retained checkpoint was assessed on the unseen test set it achieved an accuracy of 0.9082 and a macro-averaged F1 score of 0.9222. The corresponding confusion matrix indicates balanced performance across negative, neutral and positive classes with no systematic misclassification pattern (Fig. 7).

**Confusion Matrix**

|  | Negative | Neutral | Positive |
|---|---|---|---|
| **Negative** | 54 | 0 | 7 |
| **Neutral** | 2 | 74 | 3 |
| **Positive** | 2 | 3 | 73 |

**Fig. 7.** Matrix that shows correlation between predicted labels by the model and the true labels among the test dataset.

## 5.3 Comparative Analysis

Compared with previously reported baselines of the models operating on the same SEED, SEED-FRA, and SEED-GER repositories, the proposed by our study model (CNN-Transformer) outperforms a support-vector machine (SVM) (0.8665 accuracy), a deep neural network (DNN) (0.8608 accuracy) and logistic regression (LM) (0.8270 accuracy) while relying only five electrodes (Table 2).

|  | CNN-Transformer | SVM (System Vector Machine) | DNN (Deep Neural Network) | LR (Linear Regression) |
|---|---|---|---|---|
| **Test Accuracy** | 0.9082 | 0.8665 | 0.8608 | 0.8270 |
| **Number of EEG electrodes** | 5 | 12 | 62 | 62 |

**Table 1.** Table showing the difference between test accuracy between different model types.

## 6 Discussion

The current study demonstrates the successful development and implementation of a Transformer-based Convolutional Neural Network (Trans-former-

CNN) for emotion classification using EEG signals, achieving higher classification accuracy compared to previously utilized architectures, such as SVM, DNN, LR. By leveraging multicultural datasets (SEED, SEED-FRA, SEED-GER) and focusing on five electrodes mapped to the EMOTIV Insight device, the model not only outperforms previous benchmarks but also provides a scalable and consumer-accessible approach to emotion recognition. These findings contribute to the growing literature emphasizing the superiority of attention-based models in EEG decoding.

In addition to outperforming prior methods, the proposed model bridges gaps between affective neuroscience, deep learning, and real-world applications. Its use of a reduced-grade EEG headset highlights its accessibility beyond laboratory environments, enabling potential integration into smart homes, wearable technologies, and real-time digital experiences. This has profound implications not only for media consumption and personalization, but also for digital well-being, mental health support, and passive emotion monitoring.

Ethical considerations also emerge as a vital topic of discussion. As EEG data represents intimate neurological activity, ensuring informed consent, data security, and transparency in decision-making processes will be paramount, especially when deploying such systems in therapeutic, educational, or commercial settings.

The necessity of explainable AI (XAI) becomes particularly evident when emotional states are used to trigger behavioral interventions or guide content suggestions, especially in health, education, or therapeutic settings. In such domains, it is not sufficient for a system to simply be accurate users, clinicians, and stakeholders must understand why a particular emotional state was inferred and how it led to a specific recommendation or action. Lack of transparency may not only undermine user trust but could also lead to ethical concerns, misinterpretation, or unintended psychological effects. Future iterations of emotion-aware systems should integrate interpretability techniques, such as attention weight visualization, saliency maps, or model-agnostic approaches like SHAP, to offer insight into the decision-making process while maintaining high predictive performance.

## 7 Limitations

While the results are promising, several limitations should be acknowledged. First, the use of pre-recorded emotional stimuli may not fully replicate the complexity and variability of real-world emotional experiences. Second, although multicultural datasets were used, the representation of participants is still limited to specific linguistic and cultural groups, potentially restricting the generalizability of the model's findings across broader global populations. Third, the current study was conducted in an offline setting, meaning the model's robustness under real-time conditions, including dynamic noise, user movement, and attention variability, remains untested.

Another consideration is the fact that emotions fluctuate over time, even during the viewing of a single clip, but the model uses aggregated features over fixed windows. This may oversimplify temporal dynamics of affect. Furthermore, for now, the model does not account for inter-individual baseline differences in EEG activity, which may affect classification accuracy. Including a personal baseline calibration could improve model sensitivity.

Finally, despite employing Transformer-CNN architectures known for high performance, the black-box nature of deep learning introduces challenges in interpretability and explainability, which are particularly critical in clinical or therapeutic contexts.

## 8      Future Directions

Future work should extend this model to real-time environments using live EEG streaming to evaluate temporal adaptability and robustness against signal noise. Addressing cultural variability remains a priority, and the expansion of EEG datasets to include underrepresented populations will support the development of more inclusive, culturally sensitive emotion recognition systems.

Future iterations of the recommendation engine could also incorporate user personality profiles and adaptive learning mechanisms, such as reinforcement learning and user feedback loops. These features will help refine predictions and tailor suggestions dynamically to maximize emotional alignment and user satisfaction. Integration into commercial products, (smart homes, wellness platforms, streaming services) may enable seamless, passive emotional monitoring and content adjustment. Such integrations are particularly valuable in contexts where expressive behaviors (e.g., facial cues or tone of voice) are suppressed or difficult to interpret.

In neuromarketing, EEG emotion recognition can provide companies with insights into subconscious consumer responses, offering more precise targeting for product development and advertisement strategies. This application benefits from the lightweight nature of modern EEG systems, which increase ecological validity during product testing. In the healthcare sector, EEG-based emotional analysis may support users with conditions such as anxiety, depression, or PTSD through personalized digital therapeutics. These systems can serve as passive mood-regulation tools, offering emotionally congruent content or feedback that enhances well-being in daily life.

Finally, a follow-up study is already being prepared to establish a culture-specific EEG dataset from the Greek population, while also testing Transformer-based models across both cross-cultural and within-culture contexts. This forthcoming study will also explore the influence of personality traits, such as empathy and conscientiousness, on emotional responses to media. It will examine how emotional congruence or incongruence between user mood and media content affects user satisfaction and mood outcomes. These efforts aim to further advance neuro-driven, personalized content recommendation and inform best practices for inclusive, ethical, and clinically meaningful emotion-aware technology design.

**Both authors have contributed equally to the conception, design, and preparation of this manuscript.**

# References

1. Ravi D, Wong C, Deligianni F, Berthelot M, Andreu-Perez J, Lo B, et al. Deep Learning for Health Informatics. IEEE J Biomed Health Inform. 2017;21:4–21.
2. Esteva A, Robicquet A, Ramsundar B, Kuleshov V, DePristo M, Chou K, et al. A guide to deep learning in healthcare. Nat Med. 2019;25:24–9.
3. Vieira S, Pinaya WHL, Mechelli A. Using deep learning to investigate the neuroimaging corre-lates of psychiatric and neurological disorders: Methods and applications. Neurosci Biobehav Rev.
4. Topol EJ. High-performance medicine: the convergence of human and artificial intelligence. Nat Med.
5. Wei-Long Zheng, Bao-Liang Lu. Investigating Critical Frequency Bands and Channels for EEG-Based Emotion Recognition with Deep Neural Networks. IEEE Trans Auton Ment Dev. 2015;7:162.
6. Duan R-N, Zhu J-Y, Lu B-L. Differential entropy feature for EEG-based emotion classifica-tion. 2013 6th Int IEEEEMBS Conf Neural Eng NER [Internet]. San Diego, CA, USA: IEEE; 2013.
7. Schaefer A, Nils ,Frédéric, Sanchez ,Xavier, and Philippot P. Assessing the effectiveness of a large database of emotion-eliciting films: A new tool for emotion researchers. Cogn Emot. 2010;24:1153–72.
8. Liu W, Zheng W-L, Li Z, Wu S-Y, Gan L, Lu B-L. Identifying similarities and differences in emotion recognition with EEG and eye movements among Chinese, German, and French People. J Neural Eng. 2022;19:026012.
9. Alarcão SM, Fonseca MJ. Emotions recognition using EEG signals: A survey. IEEE Trans Affect Comput. 2019;10:374–93.
10. Insight - 5 Channel Wireless EEG Headset [Internet]. EMOTIV. [cited 2025 Mar 26].
11. Toa CK, Sim KS, Tan SC. Emotiv Insight with Convolutional Neural Network: Visual Atten-tion Test Classification. In: Wojtkiewicz K, Treur J, Pimenidis E, Maleszka M, editors. Adv Comput Collect Intell. Cham: Springer International Publishing; 2021. p. 348–57.
12. Harke Pratama S, Rahmadhani A, Bramana A, Oktivasari P, Handayani N, Haryanto F, et al. Signal Comparison of Developed EEG Device and Emotiv Insight Based on Brainwave Charac-teristics Analysis. J Phys Conf Ser. 2020;1505:012071.
13. Maskeliunas R, Damasevicius R, Martisius I, Vasiljevas M. Consumer-grade EEG devices: are they usable for control tasks? PeerJ. 2016;4:e1746.
14. Subasi A, Gürsoy M. EEG signal classification using PCA, ICA, LDA and support vector machines. Expert Syst Appl. 2010;37:8659–66.
15. Shukla J, Barreda-Ángeles M, Oliver J, Nandi GC, Puig D. Feature Extraction and Selection for Emotion Recognition from Electrodermal Activity. IEEE Trans Affect Comput. 2021;12:857–69.
16. Lawhern VJ, Solon AJ, Waytowich NR, Gordon SM, Hung CP, Lance BJ. EEGNet: a com-pact convolutional neural network for EEG-based brain-computer interfaces. J Neural Eng. 2018;15:056013.
17. Wang Y, Zhang L, Xia P, Wang P, Chen X, Du L, et al. EEG-Based Emotion Recognition Using a 2D CNN with Different Kernels. Bioengineering. 2022;9:231.
18. Chandra A, Tünnermann L, Löfstedt T, Gratz R. Transformer-based deep learning for pre-dicting protein properties in the life sciences. Dötsch V, editor. eLife. 2023;12:e82819.
19. Sun J, Xie J, Zhou H. EEG Classification with Transformer-Based Models. 2021 IEEE 3rd Glob Conf Life Sci Technol LifeTech [Internet]. 2021 [cited 2025 Mar 27]. p. 92–3.
20. Koelstra S, Muhl C, Soleymani M, Jong-Seok Lee, Yazdani A, Ebrahimi T, et al. DEAP: A Database for Emotion Analysis ;Using Physiological Signals. IEEE Trans Affect Comput. 2012;3:18–31.

21. Ahangaran M, Zhu H, Li R, Yin L, Jang J, Chaudhry AP, et al. DREAMER: a computational framework to evaluate readiness of datasets for machine learning. BMC Med Inform Decis Mak. 2024;24:152.
22. SEED Dataset [Internet]. [cited 2025 Mar 27].
23. Rayatdoost S, Soleymani M. CROSS-CORPUS EEG-BASED EMOTION RECOGNITION. 2018.
24. Duvinage M, Castermans T, Petieau M, Hoellinger T, Cheron G, Dutoit T. Performance of the Emotiv Epoc headset for P300-based applications. Biomed Eng OnLine. 2013;12:56.
25. Dura A, Wosiak A. EEG channel selection strategy for deep learning in emotion recognition. Procedia Comput Sci. 2021;192:2789–96.
26. Al-Nafjan A, Hosny M, Al-Ohali Y, Al-Wabil A. Review and Classification of Emotion Recognition Based on EEG Brain-Computer Interface System Research: A Systematic Review. Appl Sci. 2017;7:1239.
27. Majid Mehmood R, Du R, Lee HJ. Optimal Feature Selection and Deep Learning Ensembles Method for Emotion Recognition From Human Brain EEG Sensors. IEEE Access. 2017;5:14797–806.
28. Soleymani M, Garcia D, Jou B, Schuller B, Chang S-F, Pantic M. A survey of multimodal sentiment analysis. Image Vis Comput. 2017;65:3–14.
29. Humanized Recommender Systems: State-of-the-art and Research Issues | ACM Transactions on Interactive Intelligent Systems [Internet]. [cited 2025 Mar 27].
30. Musto C, Gemmis M de, Lops P, Narducci F, Semeraro G. Semantics and Content-Based Recommendations. In: Ricci F, Rokach L, Shapira B, editors. Recomm Syst Handb [Internet]. New York, NY: Springer US; 2022 [cited 2025 Mar 27]. p. 251–98.
31. Byrne A, Bonfiglio E, Rigby C, Edelstyn N. A systematic review of the prediction of con-sumer preference using EEG measures and machine-learning in neuromarketing research. Brain Inform. 2022;9:27.
32. Plassmann H, Venkatraman V, Huettel S, Yoon C. Consumer Neuroscience: Applications, Challenges, and Possible Solutions. J Mark Res. 2015;52:150109125622007.
33. Mitsea E, Drigas A, Skianis C. Artificial Intelligence, Immersive Technologies, and Neuro-technologies in Breathing Interventions for Mental and Emotional Health: A Systematic Review. Electronics. 2024;13:2253.
34. Khushaba R, Wise C, Kodagoda S, Louviere J, Kahn B, Townsend C. Consumer neurosci-ence: Assessing the brain response to marketing stimuli using electroencephalogram (EEG) and eye tracking. Expert Syst Appl. 2013;40:3803–12.
35. Vecchiato G, Astolfi L, De Vico Fallani F, Toppi J, Aloise F, Bez F, et al. On the Use of EEG or MEG Brain Imaging Tools in Neuromarketing Research. Comput Intell Neurosci. 2011;2011:643489.
36. Sheikh M, Qassem M, Kyriacou PA. Wearable, Environmental, and Smartphone-Based Passive Sensing for Mental Health Monitoring. Front Digit Health [Internet]. 2021 [cited 2025 Mar 28].
37. Bergin AD, Vallejos EP, Davies EB, Daley D, Ford T, Harold G, et al. Preventive digital mental health interventions for children and young people: a review of the design and reporting of research. Npj Digit Med. 2020;3:1–9.
38. Galindo-Aldana G, Montoya-Rivera LA, Esqueda-Elizondo JJ, Inzunza-Gonzalez E, Garcia-Guerrero EE, Padilla-Lopez A, et al. Mindfulness-Based Intervention Effects on EEG and Execu-tive Functions: A Systematic Review. Brain Sci. 2025;15:324.
39. Zotev V, Phillips R, Yuan H, Misaki M, Bodurka J. Self-regulation of human brain activity using simul-taneous real-time fMRI and EEG neurofeedback. NeuroImage. 2014;85 Pt 3:985–95.
40. Thibault RT, Lifshitz M, Raz A. The self-regulating brain and neurofeedback: Experimental science and clinical promise. Cortex J Devoted Study Nerv Syst Behav. 2016;74:247–61.
41. Kandel ER, Schwartz JH, Jessell TM. Essentials of Neural Science and Behavior. McGraw-Hill; 1995.
42. Neuroscience: Exploring the brain: Fourth edition | Request PDF [Internet]. ResearchGate.
43. Da Silva FL. EEG: Origin and Measurement. In: Mulert C, Lemieux L, editors. EEG - FMRI [Internet]. Berlin, Heidelberg: Springer Berlin Heidelberg; 2009 [cited 2025 Mar 26]. p. 19–38.
44. Villringer A, Mulert C, Lemieux L. Principles of multimodal functional imaging and data integration. EEG-FMRI Physiol Basis Tech Appl. New York, NY, US: Springer Science + Business Media; 2010. p. 3–17.
45. Herwig U, Satrapi P, Schönfeldt-Lecuona C. Using the International 10-20 EEG System for Positioning of Transcranial Magnetic Stimulation. Brain Topogr. 2003;16:95–9.
46. Klimesch W. EEG alpha and theta oscillations reflect cognitive and memory performance: a review and analysis. Brain Res Brain Res Rev. 1999;29:169–95.
47. Buzsáki G. Rhythms of the Brain [Internet]. Oxford University Press; 2006 [cited 2025 Mar 26].

48. Ramirez R, Vamvakousis Z. Detecting Emotion from EEG Signals Using the Emotive Epoc Device. In: Zanzotto FM, Tsumoto S, Taatgen N, Yao Y, editors. Brain Inform. Berlin, Heidel-berg: Springer; 2012. p. 175–84.
49. Liu S, Li G, Jiang S, Wu X, Hu J, Zhang D, et al. Investigating Data Cleaning Methods to Improve Performance of Brain–Computer Interfaces Based on Stereo-Electroencephalography. Front Neu-rosci [Internet]. 2021 [cited 2025 Mar 26];15.
50. Duan R-N, Zhu J-Y, Lu B-L. Differential entropy feature for EEG-based emotion classifica-tion. 2013 6th Int IEEEEMBS Conf Neural Eng NER [Internet]. 2013 [cited 2025 Mar 26]. p. 81–4.
51. Shi L-C, Jiao Y-Y, Lu B-L. Differential entropy feature for EEG-based vigilance estimation. 2013 35th Annu Int Conf IEEE Eng Med Biol Soc EMBC [Internet]. 2013 [cited 2025 Mar 26]. p. 6627.
52. Bashivan P, Rish I, Yeasin M, Codella N. Learning Representations from EEG with Deep Recurrent-Convolutional Neural Networks [Internet]. arXiv; 2016 [cited 2025 Mar 26].
53. Chiarelli AM, Croce P, Merla A, Zappasodi F. Deep learning for hybrid EEG-fNIRS brain-computer interface: application to motor imagery classification. J Neural Eng. 2018;15:036028.
54. Ekman P, Friesen WV. Constants across cultures in the face and emotion. J Pers Soc Psy-chol. 1971;17:124–9.
55. Scherer K, Schorr A, Johnstone T. Appraisal Processes in Emotion: Theory, Methods, Re-search. 2001.
56. Safdar S, Friedlmeier W, Matsumoto D, Yoo SH, Kwantes CT, Kakai H, et al. Variations of emotional display rules within and across cultures: A comparison between Canada, USA, and Japan. Can J Behav Sci Rev Can Sci Comport. 2009;41:1–10.
57. Cordaro DT, Sun R, Keltner D, Kamble S, Huddar N, McNeil G. Universals and cultural variations in 22 emotional expressions across five cultures. Emotion. 2018;18:75–93.
58. Leff J. The cross-cultural study of emotions. Cult Med Psychiatry. 1977;1:317–50.
59. Sneddon I, McKeown G, McRorie M, Vukicevic T. Cross-Cultural Patterns in Dynamic Ratings of Pos-itive and Negative Natural Emotional Behaviour. PLOS ONE. 2011;6:e14679.
60. Hampton RS, Varnum MEW. The cultural neuroscience of emotion regulation. Cult Brain. 2018;
61. Matsumoto D, Yoo SH, Nakagawa S. Culture, emotion regulation, and adjustment. J Pers Soc Psychol. 2008;94:925–37.
62. Lin Y-P, Wang C-H, Jung T-P, Wu T-L, Jeng S-K, Duann J-R, et al. EEG-based emotion recognition in music listening. IEEE Trans Biomed Eng. 2010;57:1798–806.
63. Han S, Northoff G. Culture-sensitive neural substrates of human cognition: a transcultural neuroimaging approach. Nat Rev Neurosci. 2008;9:646–54.
64. Chikazoe J, Lee DH, Kriegeskorte N, Anderson AK. Population coding of affect across stimuli, modali-ties and individuals. Nat Neurosci. 2014;17:1114–22.
65. Mushtaq F, Ibáñez A. Electroencephalography (EEG) and the Quest for an Inclusive and Global Neuro-science. Eur J Neurosci. 2025;61:e70078.
66. Hu X, Chen J, Wang F, Zhang D. Ten challenges for EEG-based affective computing. Brain Sci Adv. 2019;5:1–20.
67. Pei G, Li H, Lu Y, Wang Y, Hua S, Li T. Affective Computing: Recent Advances, Challeng-es, and Future Trends. Intell Comput. 2024;3:0076.
68. Zheng W-L, Zhu J-Y, Lu B-L. Identifying Stable Patterns over Time for Emotion Recogni-tion from EEG. IEEE Trans Affect Comput. 2019;10:417–29.
69. Rentfrow PJ, Gosling SD. The do re mi's of everyday life: The structure and personali-ty correlates of music preferences. J Pers Soc Psychol. 2003;84:1236.
70. Hirsh JB, Kang SK, Bodenhausen GV. Personalized persuasion: tailoring persuasive appeals to recipi-ents' personality traits. Psychol Sci. 2012;23:578–81.
71. Huizinga D, Elliott DS. Reassessing the reliability and validity of self-report delinquency measures. J Quant Criminol. 1986;2:293–327.
72. Brouwer A-M, Hogervorst MA. A new paradigm to induce mental stress: the Sing-a-Song Stress Test (SSST). Front Neurosci [Internet]. 2014 [cited 2025 Mar 26];8.
73. Henry JC. Electroencephalography: Basic Principles, Clinical Applications, and Related Fields, Fifth Edition. Neurology. 2006;67:2092-2092-a.
74. Ouyang D, Yuan Y, Li G, Guo Z. The Effect of Time Window Length on EEG-Based Emo-tion Recog-nition. Sensors. 2022;22:4939.